\documentclass[preprint,12pt]{elsarticle}

\usepackage{amssymb}

\journal{Astroparticle Physics}

\begin{document}

\begin{frontmatter}



\title{Discrepancies in the Monte Carlo simulations of propagation of ultra-high energy cosmic-ray photons in the geomagnetic field}

\author[a1]{P. Homola\corref{cor1}}
\ead{Piotr.Homola@ifj.edu.pl}
\author[a1]{M. Rygielski}
\cortext[cor1]{Corresponding author: Tel.: +48 12 6628348; fax: +48 12 6628012.}
\address[a1]{H.~Niewodnicza\'nski Institute of Nuclear Physics, Polish Academy of Sciences, ul.~Radzikowskiego 152, 31-342 Krak\'ow, Poland}
\begin{abstract}

The discrepancies in the results produced by the two most commonly used Monte Carlo programs for simulation of propagation of ultra-high energy cosmic ray photons 
in the presence of the geomagnetic field are presented. 
Although photons have not yet been discovered in the cosmic ray flux at highest energies, the capabilities of the present cosmic ray detectors make their discovery 
possible, according to the predictions of conventional models, within the next few years. It is therefore necessary to have a reliable and well maintained software 
for relevant simulations. The results of this paper are important for simulations of propagation of photons at energies above 10$^{19}$~eV. Photons of such high 
energies might interact with the geomagnetic field giving rise to a cascade of particles even above the atmosphere. This effect is called a ``preshower effect''. 
The preshower effect is important for air shower evolution and has to be accounted for in full Monte Carlo simulations of propagation of highest energy cosmic-ray photons.
In this paper we compare the two most frequently used Monte Carlo codes for preshower simulations: PRESHOWER, used as a stand-alone program or as a part of CORSIKA,
 and MaGICS, used as a part of AIRES.

\end{abstract}


\end{frontmatter}

\section{Introduction}
The composition of ultra-high energy cosmic rays (UHECR) has been unknown since the last half century. 
Recent results are uncertain about whether the flux of cosmic rays at highest energies is dominated by protons, heavier nuclei or whether it is composed 
by a mixture of nuclei of different masses (see e.g. Ref.~\cite{icrc-auger-composition}). Serious candidates for primary cosmic rays are also photons 
and neutrinos which are expected at small rates \cite{auger-phot-09,auger-neutr-09,icrc-auger-photons}. This paper is focused on UHE photons as cosmic rays.

Small fractions of photons in the flux of UHECR are predicted by the GZK mechanism \cite{gzk} describing interactions of charged cosmic ray nuclei with 
cosmic microwave background radiation. As a result of these interactions the energy of primary nuclei is reduced and high energy photons and neutrinos 
appear as products. If the GZK predictions are true, these photons and neutrinos should be seen at low rates in the cosmic ray flux. 
With the capabilities of present cosmic ray detectors, e.g. the Auger Observatory \cite{auger}, these ``guaranteed'' photon rates 
are likely to be verified within the next few years. An identification of photons in cosmic rays would be a great scientific achievement 
and it would open a new observation window to the Universe. On the other hand, the lack of the GZK photons in the cosmic 
ray flux would challenge our present understanding of cosmic-ray physics, and it would be even a more interesting observation. 
Thus the capability to identify photons among UHE cosmic rays is very important. A full readiness for observation of GZK photons requires a 
reliable Monte Carlo simulation code that treats not only photon-induced extensive air showers (EAS) but simulates also the propagation of 
photons before they enter the atmosphere.

The simulation of photons above the atmosphere is important because of the ``preshower effect'' \cite{preshw-mcbreen} that may occur when a 
photon traverses a region where the geomagnetic field component transverse to the photon trajectory is particularly strong. As described e.g. 
in Refs.~\cite{cpc1,phot-rev}, high energy photons in the presence of magnetic field may convert into e$^{+/-}$ pair and the newly created 
electrons emit bremsstrahlung photons, which again may convert into e$^{+/-}$ if their energy is high enough. As a result of these interactions, 
instead of a high energy photon, a bunch of particles of lower energies, so-called a preshower, reaches the atmosphere. Obviously, the occurrence 
of the preshower effect has an important impact on air shower development and changes many shower observables. 

This paper is a contribution towards the development and maintenance of a reliable codes for simulation of ultra-high energy photons 
before they enter the Earth's atmosphere. We compare the results produced by the two most commonly used Monte Carlo programs for the 
preshower effect simulation: PRESHOWER \cite{cpc1}, which can be used as a stand-alone program or as a part of CORSIKA air shower 
simulation package \cite{corsika}, and MaGICS~v.~1.4.0, used as a part of AIRES package \cite{aires}. PRESHOWER and MaGICS are the 
only programs of their class with open source codes and they come along with the most frequently used software for Monte Carlo 
cosmic-ray shower simulations. As the level of convergence of the results that can be obtained with the two programs has been found 
to be unsatisfactory, we have to make it clear to the community involved in simulations of ultra-high energy photon simulations. 

Descriptions of both programs as well as the formulation of the physical processes behind the preshower effect can be found elsewhere. 
In case of PRESHOWER, the reader is referred to Ref.~\cite{cpc1}, the MaGICS code is described at AIRES website (\cite{aires}), 
and the preshower effect is described and discussed e.g. in Refs.~\cite{cpc1,phot-rev}). 

It has to be mentioned that the PRESHOWER results were checked (see Ref.~\cite{cpc1}), to be in a good general agreement with previous 
studies: \cite{preshw-mcbreen,karakula,stanev,billoir,bedn,vankov1,vankov2}. However, the results presented in all the mentioned 
publications were obtained with the codes unavailable publicly, which made a detailed comparison with the PRESHOWER results unfeasible. 

There are three important aspects of preshower simulations and in all of them the discrepancies between PRESHOWER and 
MaGICS have been found\footnote{There is a new version of PRESHOWER which has just been released (\cite{cpc2}). 
The results of the new version, including major improvements of the algorithm and the 
update of the geomagnetic field model, are fully consistent with the first release of PRESHOWER.}. The first aspect, discussed in 
Section \ref{sec3}, is the geomagnetic field model and the calculation of the field component transverse to the primary particle motion. 
The differences in the second aspect, concerning the magnetic pair production by primary photons, 
are analyzed in Section \ref{sec4}, and the discrepancies of the third kind, 
related to the bremsstrahlung of preshower electrons, are presented in Section \ref{sec5}. As using both PRESHOWER and MaGICS to plot Figures of
Sections \ref{sec3}, \ref{sec4} and \ref{sec5} is not straightforward, a short explanation of technical details was added in the Appendix.

The paper is concluded in Section \ref{sec6}. \\

\section{The geomagnetic field and its component transverse to particle motion}
\label{sec3}
\begin{figure}
\begin{center}
\includegraphics[width=1.0\textwidth]{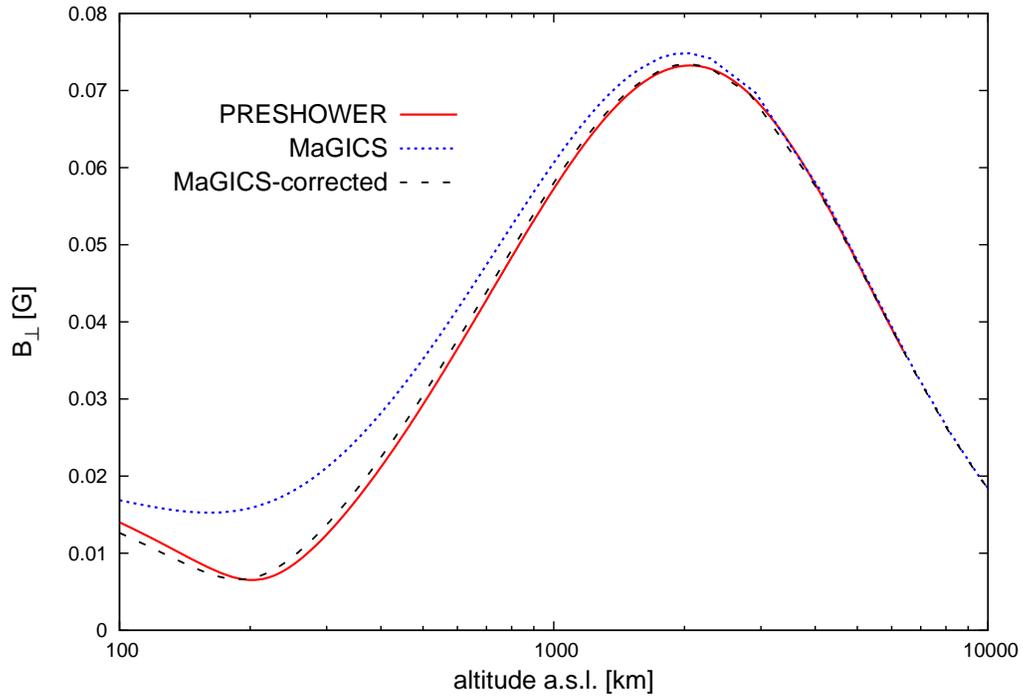}
\end{center}
\caption {Comparison of $B_\bot(r)$ for the IGRF models used in PRESHOWER (IGRF-8, solid red line), MaGICS (IGRF-10, dotted blue line) and MaGICS with applied 
corrections (dashed black line). The presented plots were obtained for the location of the Pierre Auger Observatory in Argentina and an arbitrarily selected 
trajectory with zenithal angle $\theta$=60$^\circ$ coming from geographical North.}
\label{b-pre-mag}
\end{figure}

In both PRESHOWER and MaGICS, the IGRF magnetic field model is used: in PRESHOWER -- IGRF-8 and in MaGICS -- IGRF-10. IGRF-10 is the model more 
recent than IGRF-8\footnote{The most recent version, implemented in the new release of PRESHOWER, is IGRF-11 \cite{igrf11}}. 
The differences between the models concern mainly the limit for future interpolations. 
The geomagnetic field components obtained with these two versions should agree very well for a date within the usability range of both models. Simple tests 
performed using the models implemented in PRESHOWER and MaGICS and the online geomagnetic field calculator with implemented IGRF-11 \cite{igrf-online} show a
 very good agreement of the geomagnetic field components. There is, however, a clear discrepancy in the geomagnetic field component transverse to the particle 
motion ($B_\bot$) computed by PRESHOWER and MaGICS. Figure \ref{b-pre-mag} shows an example of how $B_\bot$ changes with the altitude above the sea level for a 
primary cosmic ray coming to the Pierre Auger Observatory in Argentina from geographical North at a zenithal angle $60^\circ$ . 
The plots were obtained with PRESHOWER (solid red line), MaGICS (dotted blue line) and MaGICS with a correction described below (dashed black line). 
The differences reach a factor 2 for lower altitudes. Since the geomagnetic field components in the two programs are in a good agreement it is clear 
that the difference between $B_\bot$(PRESHOWER) and $B_\bot$(MaGICS) must come from the procedures computing $B_\bot$. After a short investigation a 
bug was found in one of the routines performing rotations in MaGICS. The existence of this bug has been verified in a simple test with an inclined trajectory 
coming from the geographic North. In a separate routine MaGICS calculates the geographic latitude and longitude of a point along the primary photon trajectory. 
These coordinates are the input for the procedure computing the geomagnetic field components. If the calculation of the geographical coordinates performed in 
MaGICS is correct, then the longitude of each of the points along the mentioned trajectory should remain unchanged. Surprisingly, the MaGICS longitude varies 
along the tested trajectory. A simple correction of a sign in one of the rotations performed by MaGICS fixed the problem. The transverse geomagnetic field 
profile produced by MaGICS with the above correction is shown in Fig.~\ref{b-pre-mag} and its agreement with the PRESHOWER curve is very good. The fixed 
procedure is used not only in the calculation of the geographic coordinates but also to obtain pair production and bremsstrahlung probabilities in MaGICS. 
Is is therefore a reasonable approach to continue the comparison of the other results using the MaGICS version including the correction described above. 
Throughout this paper we will refer to this corrected version as to ``MaGICS-corrected''.

\section{Magnetic pair production of ultra-high energy photons}
\label{sec4}

\begin{figure}
\begin{center}
\includegraphics[width=1.0\textwidth]{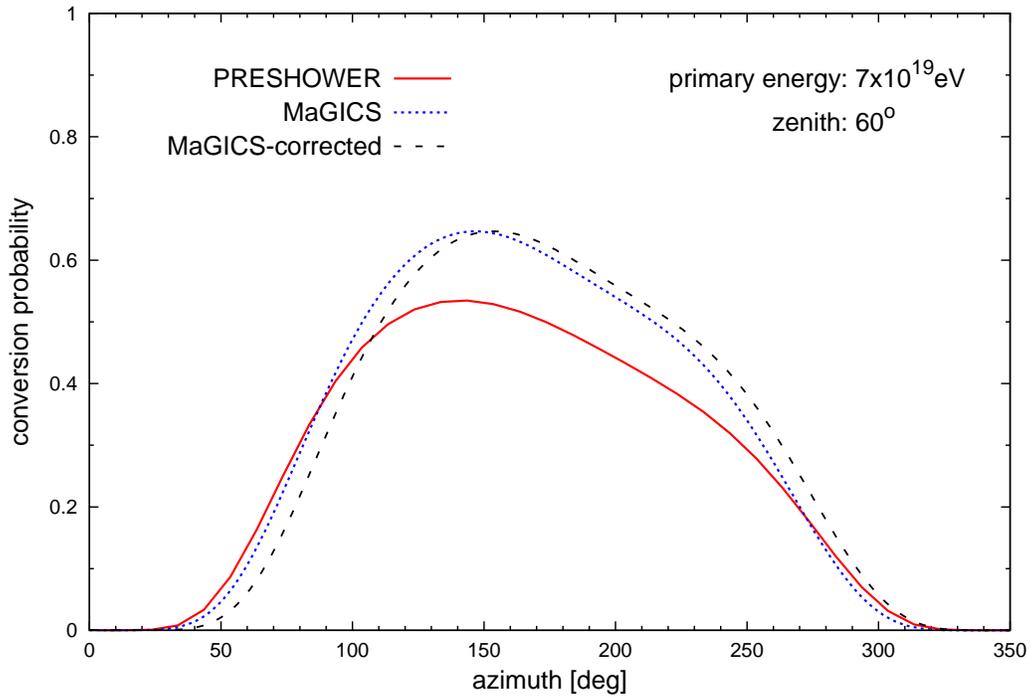}
\end{center}
\caption {Total probability of $\gamma$ conversion for primary energy of 7$\times$10$^{19}$~eV and different arrival directions as computed by PRESHOWER 
(solid red line) MaGICS (dotted blue line) and MaGICS-corrected (dashed black line). Computations have been performed for magnetic conditions at Pierre Auger 
Observatory in Argentina. The azimuth increases counterclockwise and its value of $0^\circ$ means the shower arriving from geographic North.}
\label{maps-pre-mag}
\end{figure}

Another important parameter calculated by the preshower programs is the probability of pair production by ultra-high energy photons in the presence of the
 geomagnetic field. This parameter will be hereafter referred to as gamma conversion probability or $p_{conv}$. $p_{conv}$ is a crucial parameter in photon 
shower simulation as it determines the occurrence of preshowers above the atmosphere. If an ultra-high energy cosmic ray photon initiates a preshower, the 
resultant air shower reaches its maximum much earlier than a shower developing without a preshower. It is therefore desirable that the available simulations 
give similar approximations of this crucial parameter. Unfortunately this is not the case. Neither MaGICS nor MaGICS-corrected give $p_{conv}$ in agreement 
with PRESHOWER. The differences in $p_{conv}$ reach 20\% and depend on the arrival direction. In Fig.~\ref{maps-pre-mag} we show an example of $p_{conv}$ 
dependence on the arrival direction for PRESHOWER, MaGICS and MaGICS-corrected. 
The plots were obtained for a primary photon of energy 7$\times$10$^{19}$~eV arriving at the Pierre Auger Obsrevatory at a fixed zenithal angle of 60$^\circ$ 
and varying azimuth. In the frame of reference used to prepare Fig.~\ref{maps-pre-mag}, the azimuth increases counterclockwise and its value of $0^\circ$ means 
the shower arriving from geographic North. It is clear that the results from MaGICS and PRESHOWER seriously disagree, even after a correction described earlier. 
The source of this disagreement remains unknown and requires a deeper analysis. 

\section{Magnetic bremsstrahlung of preshower electrons}
\label{sec5}

\begin{figure}
\begin{center}
\includegraphics[width=1.0\textwidth]{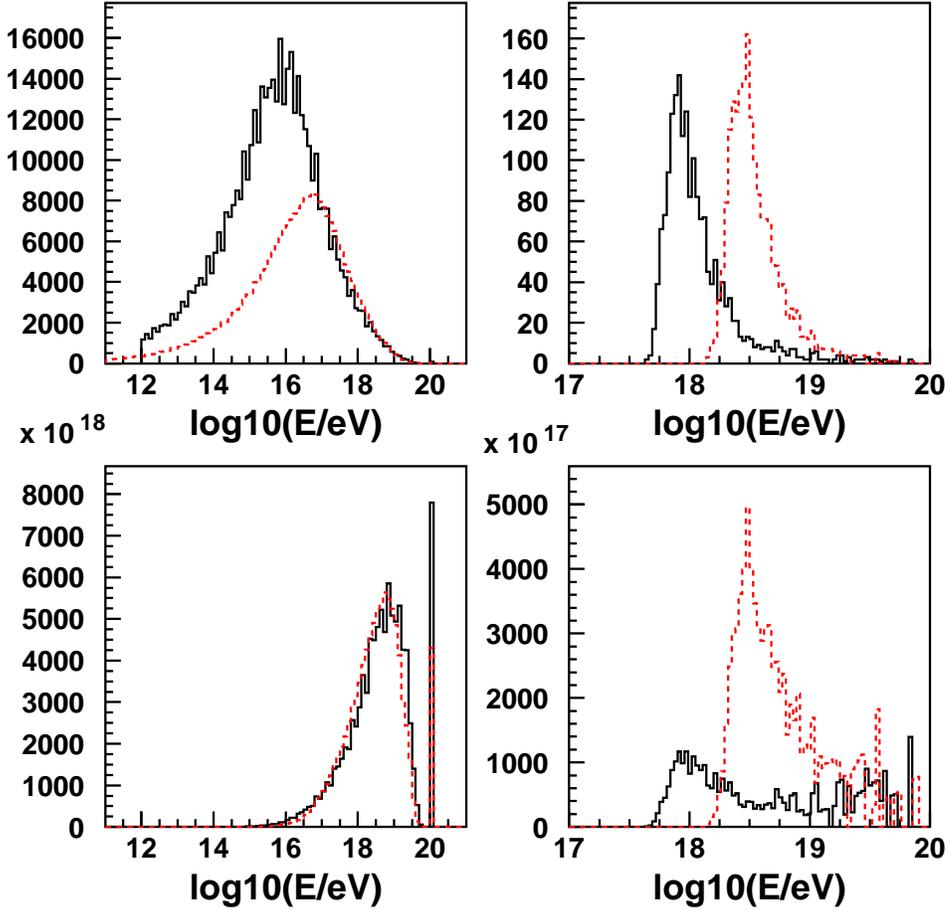}
\end{center}
\caption {Energy distribution of photons (top left) and electrons (top right) in 1000 simulation runs with
$10^{20}$~eV primary photons arriving at the Pierre Auger Observatory from geographical South at a zenithal angle of $60^\circ$. The spectra weighted by 
energy are plotted in the bottom panel. The black solid histograms were obtained with PRESHOWER and the dashed red ones with MaGICS-corrected. 
The histograms obtained with original MaGICS are not shown to keep the plots clear. It was checked that the difference between the MaGICS and MaGICS-corrected 
histograms is very small and would be hardly visible in the scale suitable for comparison to PRESHOWER histograms. The histograms shown include all the
 particles and no normalization was applied. The height of the bin at log10(E/eV)=20 of the MaGICS-corrected histogram (dashed red) in the lower left plot, 
 43$\times10^{20}$~eV, might be hardly readable in a black-and-white version of the paper.}
\label{fig6-pre-mag}
\end{figure}

\begin{figure}
\begin{center}
\includegraphics[width=1.0\textwidth]{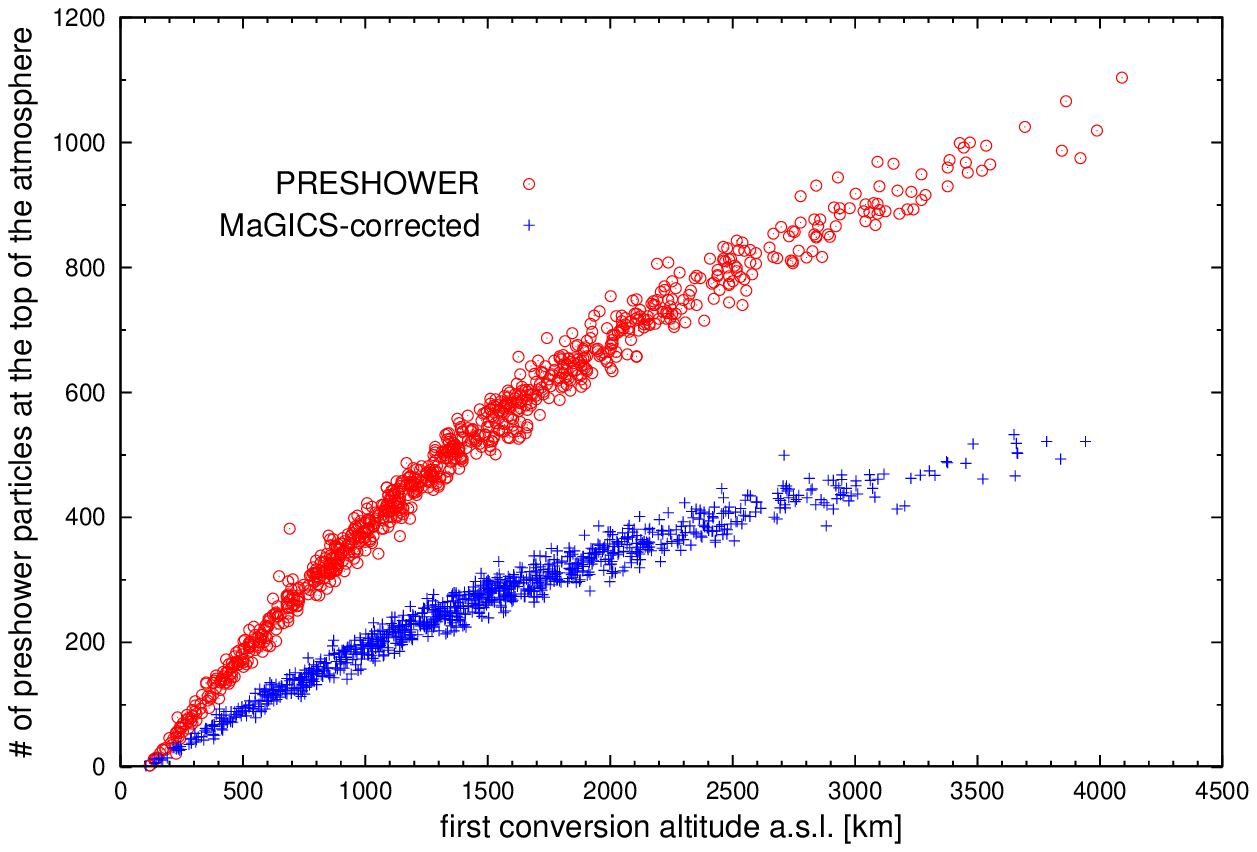}
\end{center}
\caption {Number of preshower particles at the top of the atmosphere (100~km a.s.l.) for different altitudes of
primary $\gamma$ conversion simulated with PRESHOWER (red circles) and MaGICS-corrected (blue crosses). All the simulated preshowers were initiated by
 photons of energies $10^{20}$ eV arriving from the ``strong field direction''. The results obtained with original MaGICS were checked to be very close 
 to those obtained with MaGICS-corrected and they are not shown here to keep the plots clear.}
\label{npart-pre-mag}
\end{figure}

\begin{figure}
\begin{center}
\includegraphics[width=1.0\textwidth]{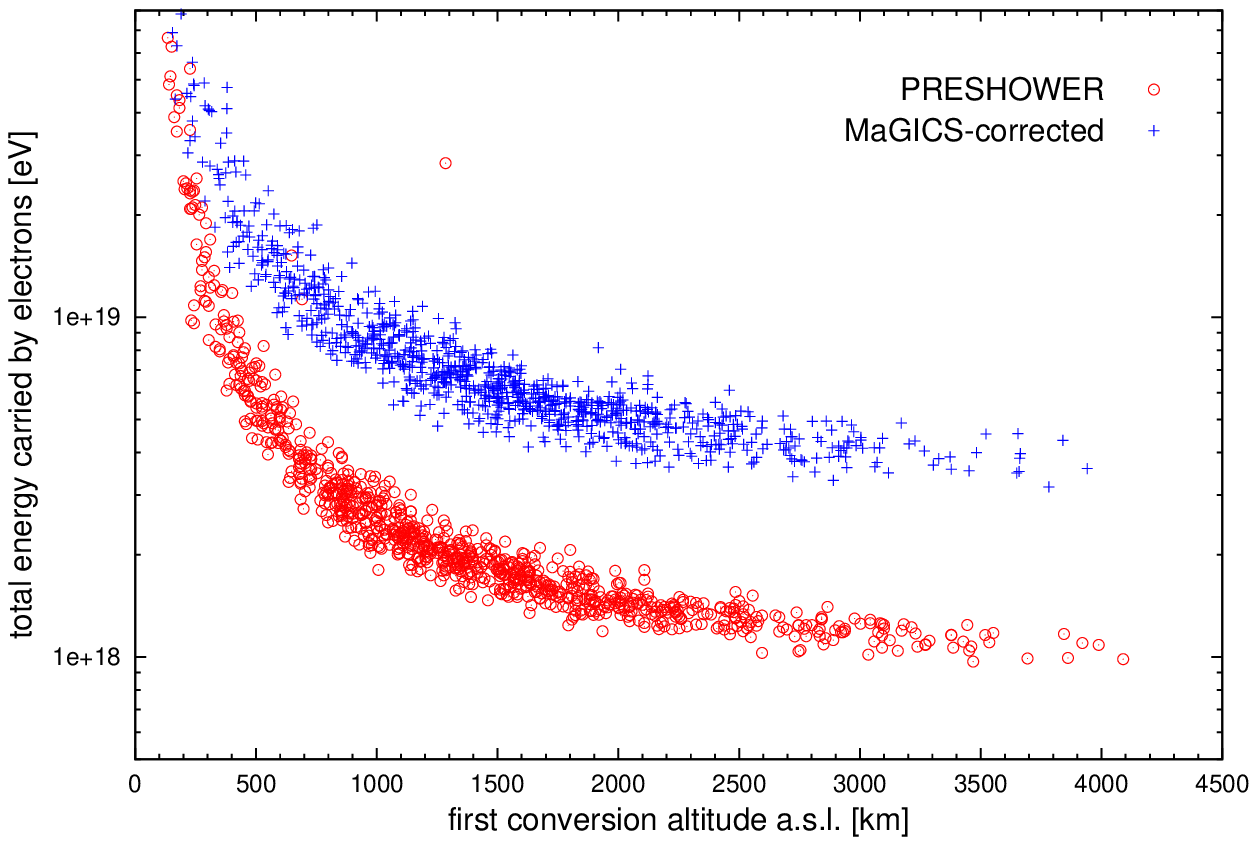}
\end{center}
\caption {Energy carried by preshower electrons at the top of the atmosphere (100~km~a.s.l.) vs.
the altitude of the primary $\gamma$ conversion for a primary photon energy of
$10^{20}$~eV arriving from the ``strong field direction''. The simulations were performed with PRESHOWER (red circles) and MaGICS-corrected (blue crosses). 
The points in excess of the general trend are the cases where one of the bremsstrahlung photons converted into an electron pair which increased the total 
energy carried by electrons. The results obtained with original MaGICS were checked to be very close to those obtained with MaGICS-corrected and they are 
not shown here to keep the plots clear.
}
\label{ee-pre-mag}
\end{figure}

The third important difference between MaGICS and PRESHOWER is in the probability of magnetic bremsstrahlung emission by preshower electrons, hereafter $p_{brem}$. 
Large $p_{brem}$, or in other words more efficient bremsstrahlung emission, results in splitting the primary energy into more particles of lower energies than in 
case of smaller $p_{brem}$. Fig.~\ref{fig6-pre-mag} shows some preshower statistics for an example of primary photon of energy $10^{20}$~eV arriving from 
geographical South at the Pierre Auger Observatory at a zenithal angle of $60^{\circ}$. Along this direction the transverse component of the geomagnetic 
field is particularly strong.
 We will call this direction a ``strong field direction''. The histograms shown include all the particles and no normalization was applied. A comparison is 
 made between the simulations obtained with PRESHOWER (solid black histograms) and MaGICS-corrected 
 (dashed red histograms)\footnote{It was checked that the original version of MaGICS gives results which are very similar to MaGICS-corrected, therefore 
 the histograms obtained with original MaGICS have not been shown in Fig.~\ref{fig6-pre-mag} to keep the plots clear.}. 1000 simulation runs have been 
 performed with each of the programs for a photon primary of energy $10^{20}$~eV arriving at the Pierre Auger Observatory from geographical South at a 
 zenithal angle of $60^\circ$. For such an arrival direction and primary energy the conversion probability $p_{conv}$ turns out to be higher in MaGICS (0.96) 
 than in PRESHOWER (0.92). As a result, more photons simulated with MaGICS happened to undergo the pair production process and initiated a preshower than it was 
 in the case of PRESHOWER simulations. Despite the larger total number of preshowers resulting from MaGICS-corrected simulations, the energy spectrum of photons 
 in the top left plot of Fig.~\ref{fig6-pre-mag} shows that the total number of secondary photons in preshowers generated by MaGICS-corrected is significantly smaller
  than in the case of PRESHOWER simulations. The significantly smaller total number of photons in MaGICS-corrected simulations shows then a serious inconsistence between
   the analyzed programs. The smaller number of photons in MaGICS-corrected is consistent with higher energies of electrons simulated by this program, which can be seen
    in the plots in the bottom row of Fig.~\ref{fig6-pre-mag}. The total number of preshower particles and total energy carried by electrons in PRESHOWER and 
    MaGICS-corrected are compared directly in Figs.~\ref{npart-pre-mag} and~\ref{ee-pre-mag}.
The results obtained with original MaGICS were checked to be very close to those obtained with MaGICS-corrected and they are not shown here to keep the plots clear.
The discrepancy between the results of PRESHOWER and MaGICS-corrected is significant. Preshower electrons in MaGICS emit significantly fewer bremsstrahlung 
photons than in PRESHOWER simulations. This must have implications on air shower development: an air shower simulated with increased $p_{brem}$ develops earlier in the atmosphere.
The source of this discrepancy, as well as the reason for the disagreement in $p_{conv}$, remains unknown and requires a further study.

\section{Summary}
\label{sec6}
We have compared the simulation results of the two most popular open source codes for preshower simulations, PRESHOWER and MaGICS. 
A significant disagreement was found between the main physical quantities derived by the two programs: 
the component of the geomagnetic field transverse to the primary trajectory ($B_\bot$), 
the probability of primary gamma conversion into an electron-positron pair ($p_{conv}$), 
and the probability of bremsstrahlung emission of preshower electrons ($p_{brem}$). 
The source of the disagreement was found only in the case of $B_\bot$ computation. 
After fixing the coding mistake found in MaGICS a good agreement between the $B_\bot$ values 
obtained with the two programs has been achieved. The sources of the other two discrepancies remain unknown and require a further study. 
The discrepancies between the results of PRESHOWER and MaGICS are meaningful for simulations of air shower development.
Taking into account that an air shower is simulated as a simple superposition of air showers induced by each of the preshower particles, 
one can assume that the depth of shower maximum, $X_{max}$, is determined mostly by the highest energy particle in a preshower that reaches 
the top layers of the atmosphere. This highest energy particle could be a photon or an electron, depending on the preshower development and the 
geomagnetic conditions. If it is a photon, then based on the results presented in Fig. \ref{fig6-pre-mag} one would 
expect only a minor difference in $X_{max}$, as photons of high energies are similarly numerous in both PRESHOWER and MaGICS.
But if an air shower development is determined mostly by a preshower electron, then the resulting $X_{max}$ could be even a few tens of g/cm$^{2}$ deeper 
if we use a preshower computed by MaGICS than in case of using PRESHOWER. This is because of the difference in electron energy calculated by the 
two programs (see Fig.~\ref{ee-pre-mag}) and a known direct relation between the energy of primary particles and $X_{max}$: a higher primary energy results 
in a deeper $X_{max}$ of an air shower. 
A quantification of the influence of the discrepancies between PRESHOWER and MaGICS on air shower simulations 
would require a separate analysis. 
Nevertheless, it is clear already now that reaching an agreement between results derived with PRESHOWER and MaGICS would be an important step towards a better 
reliability of ultra-high energy photon shower simulations. 

\section*{Acknowledgements}

We thank Henryk Wilczy\'nski for valuable discussions.\\

This work was partially supported by the Polish Ministry of Science and Higher Education under grant No.~N~N202~2072~38.

\appendix

\section{Reading variables from PRESHOWER and MaGICS}

A thorough comparison of the results produced by PRESHOWER and MaGICS requires a few manual interventions within the codes of the programs. 
These interventions are needed to read all the variables of interest and not all of these variables are delivered at the outputs. 
Additional difficulty is that MaGICS is not a stand-alone program and can be used only within its mother program AIRES. 
Although we are not aware of any fundamental reasons for which MaGICS could not be detached from AIRES (with some coding effort),
we gave up this idea to minimize the risk of errors. With this approach the computing has increased but it was still acceptable.

In order to obtain the plots presented throughout the paper we extracted the following variables:
\begin{itemize}
 \item $B_{bot}$[G] (Fig. \ref{b-pre-mag}): \texttt{B\_tr} (P); \texttt{bperp}$\times$10$^{-5}$ (M)
 \item altitude a.s.l.[km] (Fig. \ref{b-pre-mag}): \texttt{r\_curr}$\times 10^{-5}$-6371 (P); \\
\texttt{this\_particle->altitude}$\times$10$^{-3}$ (M)
 \item conversion probability (Fig. \ref{maps-pre-mag}): \texttt{p\_conv} (P); \\
function \texttt{conversion\_probability()} (M)
 \item azimuth [deg] (Fig. \ref{maps-pre-mag}): \texttt{the\_deg} (P); \\
\texttt{atan2(shower\_axis[1],shower\_axis[0])}$\times$180/$\Pi$ (M)
 \item particle type (Figs. \ref{fig6-pre-mag}, \ref{npart-pre-mag} and \ref{ee-pre-mag}): output file \texttt{part\_out.dat} (P);\\
 \texttt{this\_particle->particleID} (M)
 \item particle energy (Figs. \ref{fig6-pre-mag}, \ref{npart-pre-mag} and \ref{ee-pre-mag}): output file \texttt{part\_out.dat} (P);\\
 \texttt{this\_particle->energy} (M)
 \item primary gamma conversion altitude [km] (Figs. \ref{npart-pre-mag} and \ref{ee-pre-mag}): output file \texttt{multirun.dat} (P);
 \texttt{first\_particle->altitude}$\times$10$^{-3}$ (M)
\end{itemize}
where (P) stays for PRESHOWER and (M) for MaGICS.


\end{document}